\documentclass[aps,prl,twocolumn,showpacs,superscriptaddress,groupedaddress, final, nofootinbib]{revtex4}  
\usepackage{graphicx}  
\usepackage{dcolumn}   
\usepackage[utf8]{inputenc}
\usepackage[T1]{fontenc}
\usepackage[ngerman, english]{babel}

\usepackage{setspace} 

\usepackage[colorinlistoftodos, prependcaption, textsize=tiny]{todonotes}
\usepackage{soul}

\usepackage{blindtext}

\usepackage{amsmath, amsfonts, amsthm, amssymb} 
\usepackage{feynmp-auto}    
\usepackage{simplewick}     
\usepackage{braket}         
\usepackage{tensor}         
\usepackage{bm, bbm}            
\usepackage{mathrsfs}       

\newcommand{\diff}[1]{\text{d}{#1} \, }

\newcommand{\Z}{\ensuremath{\mathcal{Z}}}
\newcommand{\M}{\ensuremath{\mathcal{M}}}
\newcommand{\K}{\ensuremath{\mathcal{K}}}
\newcommand{\g}{\ensuremath{\gamma}}
\newcommand{\oneby}[1]{\frac{1}{#1}}

\begin{document}

\title{Sampling General N-Body Interactions with Auxiliary Fields}
\author{Christopher K\"orber}
\email{c.koerber@fz-juelich.de}
\affiliation{Institut f\"ur Kernphysik and
    Institute for Advanced Simulation   \\
    Forschungszentrum J\"ulich          \\
	52425 J\"ulich Germany}
\author{Evan Berkowitz}
\email{e.berkowitz@fz-juelich.de}
\affiliation{Institut f\"ur Kernphysik and
    Institute for Advanced Simulation   \\
    Forschungszentrum J\"ulich          \\
	52425 J\"ulich Germany}
\author{Thomas Luu}
\email{t.luu@fz-juelich.de}
\affiliation{Institut f\"ur Kernphysik and
    Institute for Advanced Simulation   \\
    Forschungszentrum J\"ulich          \\
	52425 J\"ulich Germany}
\date{\today}

\begin{abstract}
We present a general auxiliary field transformation which generates effective interactions containing all possible N-body contact terms.
The strength of the induced terms can analytically be described in terms of general coefficients associated with the transformation and thus are controllable.
This transformation provides a novel way for sampling 3- and 4-body (and higher) contact interactions non-perturbatively in lattice quantum monte-carlo simulations.
We show that our method reproduces the exact solution for a two-site quantum mechanical problem.
\end{abstract}

\pacs{pacs}
\maketitle

The introduction of auxiliary fields, e.g. via the Hubbard-Stratonovitch (HS) transformation \cite{Stratonovich1957,Hubbard1959}, as a means of linearizing the interaction term of the Hamiltonian in terms of its density operator is common practice in many areas of theoretical physics, and is particularly prevalent in condensed matter and nuclear physics \cite{gubernatis2016quantum,Kleinert:2011rb}.
Upon linearization of the theory, the problem becomes that of many particles undergoing one-body interactions with the fluctuating background auxiliary field.
In occupation number formalism, the problem becomes very similar to that of \emph{non-interacting} (amongst themselves) particles, which in some cases can be evaluated via steepest-descent methods (\emph{e.g.} 1-D Ising model).
At the very least, the transformation greatly facilitates numerical
treatment of the many-body problem \cite{Lee:2008fa,Epelbaum:2009pd,Epelbaum:2009,Carlson:2014vla}.
In path integral formalisms that involve Grassmann fields, the transformation is essential as it reduces the Lagrangian to terms bilinear in fermionic fields, which can subsequently be integrated out exactly via Grassmann gaussian integration (\emph{e.g.} bosonization of fermionic theories \cite{kopietz2008bosonization}).
The HS transformation is ideal for theories that initially have 2-body interactions (i.e. terms quadratic in the density operator).
In principle many-body interactions can be linearized through recursive application of the HS transformation, but at the cost of introducing numerous auxiliary fields.  A transformation that naturally includes 3-body interactions, for example, would be beneficial for studying ultra-cold gases of polar molecules, where 3-body (and higher) interactions can be tuned to become dominant \cite{buchler2007,Bonnes2010}.

In this Letter we detail a generalization of the HS transformation that includes two-body and all possible \emph{higher} contact interactions that involves only one auxiliary field.
The standard HS transformation can be viewed as a particular limit of this general transformation.
Our transformation, in principle, induces all possible $n$-body interactions
    $\lambda^{(n)}\hat{\rho}^n$
where $\hat{\rho}$ is the density operator.
The coefficients $\lambda^{(n)}$ are analytically determined and controllable through a set of accompanying coefficients $c_j$ that control the coupling of the density operator with the $j^{\text{th}}$ power of the auxiliary field in the linearized theory.
The numerical implementation of this transformation is trivial as the sampling of the fields can be obtained from known distributions.
For 2-flavor fermionic systems (e.g. nucleons), this transformation allows for complete control of contact interactions up to 4-body in nature.

\emph{Formalism.}\label{sec:formalism}
Consider the following integral which is intended to correspond to the interacting part of a partition function at a single space-time point
\begin{equation}
	\label{eq:Zc}
	\Z_{c, N} 
	\equiv
	\int\limits_{-\infty}^\infty \diff{\phi} P_N(\phi)
		\exp\left\{ - \sum\limits_{j=1}^{2N-1} c_j \phi^j \hat
                  \rho \right\}\ ,
\end{equation}
where $P_N(\phi)$ is the normalized probability distribution
\begin{equation}
P_N(\phi)= 	\frac{N}{ \Gamma \left( \frac{1}{2N}  \right) } e^{ - \phi^{2N}} \ .
\end{equation}
Here, $\phi$ is an auxiliary field which couples to the fermionic density operator $\hat \rho = \sum_{f} {\bar{\psi}}_{f} \psi_{f}$, where $f$ runs over the different fermion species at a given site.
Because the density operators at different spacetime points commute, we present only the derivation for a single point.

The argument of the exponential in \eqref{eq:Zc} describes interaction vertices with an incoming and outgoing fermionic field (the density operator) and from one up to $2N-1$ auxiliary fields associated with interaction strength $c_j$.
The integral is normalized such that $\Z_{c, N} \to 1$ for $c_j \to 0$.
The result of the integration over the largest exponent of $\phi$ times another polynomial in $\phi$ is given by
{\small\begin{equation}
	\int\limits_{-\infty}^\infty \diff{\phi} e^{ - \phi^{2N}} \phi^{2k}
	=
	\frac{
		\Gamma\left(\frac{1 + 2 k}{2 N}\right)
	}{N}
	\overset{N\to\infty}{\longrightarrow}
	\frac{2}{1 + 2 k}
	\, , \quad
	\forall k \in \mathbb{N}_0, N \in \mathbb{N}
	\, .
\end{equation}}
It is sufficient to only consider polynomials in even powers of $\phi$ because of the symmetry of the integration -- odd powers vanish.
Also, in order for the integral to converge, the leading exponent is not allowed to be odd in $\phi$.

In this work we identify the integral $\Z_{c,N}$ with an effective action consisting of general $2k$-fermionic field vertices
\begin{align}\label{eq:Zla}
	\Z_{\lambda}
	&\equiv
	\exp\left\{
		-\sum\limits_{k=1}^{\infty} \lambda^{(k)} \hat \rho ^k
	\right\}
	\, .
\end{align}
We expand $\Z_{c,N}$ and $\Z_\lambda$ in powers of the density operator $\hat \rho$ and systematically match order-by-order, relating the auxiliary field interactions to the induced many-body forces.
Using Fa\`{a} di Bruno's formula \cite{FdB1855} we find that
\begin{align}
	\Z_{c, N}
	&=
	\sum_{M=0}^\infty
	\Z_{c, N}^{(M)} \hat \rho^M
	\, ,    \\
    \label{eq:Zcresfull}
    	\Z_{c, N}^{(M)}
    	&=
    	\sum\limits_{k = \left \lceil{M/2}\right \rceil }^{\left \lfloor{(2N-1)M/2}\right \rfloor}
    	\frac{\Gamma\left(\frac{1 + 2k}{2 N}\right)}{ \Gamma \left( \frac{1}{2N}  \right) }
    	\sum \limits_{\vec m \in \M_{NM}^{(2k)}}
    	\prod\limits_{j=1}^{2k}
    		\left[
    			\frac{ (-c_j)^{m_j} }{m_j!}
    		\right]
\end{align}
where the sum runs over the set
{\small
\begin{align}
    \M_{NM}^{(2k)}
    =&
    \left\{
        \vec m \in \mathbbm{N}^{2k}_0 \, \middle\vert \,
        \left(\sum_{j=1}^{2k} \, j \, m_j = 2k  \right) \right. \nonumber\\
    &\left.
    \land\left(\sum_{j=1}^{2k}   \, m_j = M \right)
    \land\left(j\geq 2N \Rightarrow m_j=0\right)
    \right\}
    \, ,
\end{align}}
and
{\small
\begin{align}
	\Z_{\lambda}
	=
	\sum_{M=0}^\infty
	\Z^{(M)}_{\lambda} \hat \rho^M
	\, ,    \quad
    \label{eq:dZla}
    \Z^{(M)}_{\lambda}
    = \hspace*{-10pt}
    \sum \limits_{\vec m \in \M^{(M)}}
    \prod\limits_{k=1}^M 
    \left[
        \frac{ \left(- \lambda^{(k)} \right)^{m_k} }{m_k!} 
    \right]
\end{align}}
where the sum in $\Z_\lambda^{(M)}$ is over the set
\begin{equation}
	\M^{(M)}
	=
	\left\{
		\vec m \in \mathbbm{N}^M_0 \, \middle\vert \,
		\sum_{k=1}^M \, k \, m_k = M
	\right\}
	\, .
\end{equation}
As the highest coefficient $\lambda^{(k)}$ (in terms of $k$) in \eqref{eq:dZla} is given by $\lambda^{(M)}$ and is also linear in exactly this coefficient (because if $m_M = 1$ then $m_{j\neq M} = 0$), one can recursively determine all coefficients $\lambda^{(M)}$ for $M > 0$ by
{\small
\begin{align}
	\Z^{(M)}_{\lambda}
	&=
	-\lambda^{(M)} 
	+
	\Z^{(M)}_{\lambda}\big|_{\lambda^{(M)}\to 0}
	=
	\Z^{(M)}_{c,N}
	\, ,
	&
	\Z^{(1)}_{\lambda}\big|_{\lambda^{(1)}\to 0}
	&=
	0
	\, .
\end{align}}
Note that one can prove by induction that each coefficient $\lambda^{(M)}$ is proportional to a sum where each term is a products of $M$ fermion auxiliary-field coefficients: $\lambda^{(M)} \propto c_{j_1} \cdots c_{j_M}$ \cite{our-long-paper}.
In Tab.~\ref{tab:coeff} we show the coefficients of the induced forces up to the order of four-body forces ($M=4$) for three different choices of $N$.

\begin{table*}
    \begin{center}
{\small
    \begin{tabular}{|c|c|p{0.4\textwidth}|p{0.4\textwidth}|}
        \hline
                        &   $N=1$               &   $N=2$   
                                    &   $N=\infty$, $c_{j>3}=0$
                        \\\hline\hline
        $\lambda^{(1)}$ &   $0$                 &   $\g_{3,1} c_2$
                                    &   $\frac{1}{3}c_2$
                        \\\hline
        $\lambda^{(2)}$ &   $-\frac{1}{4}c_1^2$ &   $-\left(\oneby{8}-\frac{\g_{3,1}^2}{2}\right)c_2^2-\oneby{4}c_1c_3-\oneby{2}\g_{3,1}c_1^2-\frac{3}{8} \g_{3,1} c_3^2$
                                    &   $-\oneby{6}c_1^2 - \frac{2}{45}c_2^2 - \oneby{5} c_1 c_3 - \oneby{14}c_3^2$
                        \\\hline
        $\lambda^{(3)}$ &   $0$                 &   $\left(\oneby{8} - \frac{\g_{3,1}^2}{2}\right)c_1^2c_2 + \left(\frac{5}{32} - \frac{3\g_{3,1}^2}{8}\right) c_2c_3^2$
                                    &   $\frac{2}{45}c_1^2c_2 + \frac{8}{2835}c_2^3 + \frac{8}{105} c_1c_2c_3 + \frac{2}{63}c_2c_3^2$
                        \\
                        &                       &   $+ \oneby{2} \g_{3,1}c_1c_2c_3 + \oneby{3}\g_{3,1}^3c_2^3$
                                    &
                        \\\hline
        $\lambda^{(4)}$ &   $0$                 &   $\left(\frac{\g_{3,1}^2}{8}-\oneby{96}\right)c_1^4 - \frac{\g_{3,1}^3}{2}c_1^2c_2^2 - \left(\frac{3}{64}-\frac{3\g_{3,1}^2}{16}\right)c_1^2c_3^2$
                                    &   $\oneby{180}c_1^4 + \frac{4}{14 175}c_2^4 + \oneby{105}c_1^3c_3 - \frac{52}{10 395} c_2^2 c_3^2 - \frac{5}{7644} c_3^4$
                        \\
                        &                       &   $- \frac{\g_{3,1}}{8}c_1c_3^3 - \left(\oneby{8}-\frac{\g_{3,1}^2}{2}\right)c_1c_2^2c_3 - \frac{\g_{3,1}+3\g_{3,1}^3}{8}c_2^2c_3^2$
                                    &   $- \frac{4}{945}c_1^2c_2^2 + \frac{13}{3150} c_1^2 c_3^2 - \frac{16}{1575} c_1c_2^2c_3 - \oneby{1155} c_1c_3^3$
                        \\
                        &                       &   $ - \left(\oneby{192}-\frac{\g_{3,1}^4}{4}\right) c_2^4 - \left(\frac{15}{512} - \frac{9\g_{3,1}^2}{128}\right)c_3^4$
                                    &
                        \\
        \hline
    \end{tabular}
}
    \end{center}
    \caption{ We show the results of the matching for $N=1$, $N=2$, and $N=\infty$ with all $c_{j>3}=0$.  The first column is the Hubbard-Stratonovich case, which produces only a two-body interaction.  We repeatedly used $\Gamma(x)=(x-1)\Gamma(x-1)$ to simplify many $N=2$ coefficients, and use the shorthand $\g_{3,1}=\Gamma(3/4)/\Gamma(1/4)$.  We provide a {\it Mathematica} notebook useful for generating the $\lambda^{(M)}$ for a given $N$ in the Supplementary Material.}
    \label{tab:coeff}
\end{table*}

\emph{Numerical Results.}
\begin{figure*}[t!]
\includegraphics[width=\textwidth]{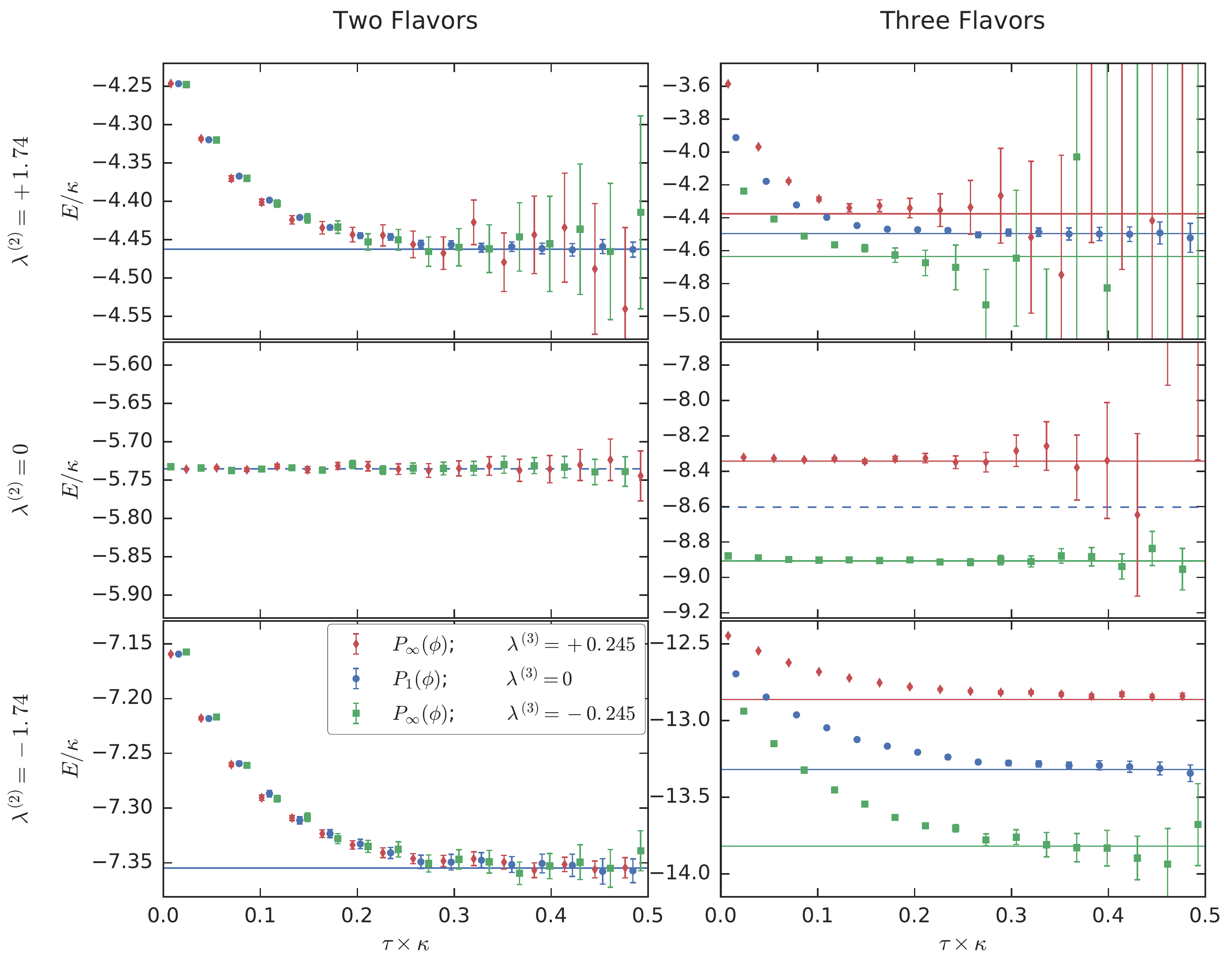}
\caption{$E(\tau)/\kappa$ for different systems.    All systems were studied with $2\times10^8$ measurements.  Some systems show a
  signal-to-noise problem and others yield an easy extraction of a
  constant plateau in the long-time-limit.  The left panels all show
  two-fermion systems and the corresponding right panels show the
  three-fermion systems with the same parameters.  In the top (middle)
  [bottom] two panels we show a system with repulsive (absent)
  [attractive] two-body forces.  In the top four panels red (blue)
  [green] points correspond to repulsive (absent) [attractive] three
  body forces, where the blue points were sampled according to the HS
  distribution $P_1$ and the other points according to $P_\infty$.
  The data in the mid two panels were generated with $P_\infty$ and
  $c_j<4$ coefficients tuned to exactly cancel the two-body force, and
  we show dashed lines for the corresponding non-interacting energies.
 }
\label{fig:spectrum}
\end{figure*}
To demonstrate the efficacy of our transformation, we consider a system of 3 different fermion species $f$ interacting on a two-site model (with sites $\{0,1\}$) with the following Hamiltonian,
{\small\begin{equation}\label{eqn:hamiltonian}
\hat H=\kappa\left(-3\sum_{f=1}^3\sum_{\langle i,j\rangle}\hat a_{f,i}\hat a^\dag_{f,j}+\lambda^{(2)}\sum_{i\in\{0,1\}} \hat \rho_i^2+\lambda^{(3)}\sum_{i\in\{0,1\}} \hat \rho_i^3\right)\ ,
\end{equation}}
where $\hat a^\dag_{f,i}$ ($\hat a_{f,i}$) is a creation (annihilation) operator for a fermion of species $f$ at site $i$ and $\hat \rho_i=\sum_f \hat a^\dag_{f,i}\hat a_{f,i}$ is the number operator at site $i$.  Here the sum over $\langle i,j\rangle \in \{ (0,1), (1,0) \}$ represents nearest neighbor hoppings between the two sites, and $\kappa$ is a dimensionful parameter that sets the dynamical scale of our problem.  As is evident in \eqref{eqn:hamiltonian}, $\lambda^{(2)}$ and $\lambda^{(3)}$ represent the size of two- and three-body interactions (relative to $\kappa$), respectively, and are dimensionless.  A positive (negative) value of $\lambda^{(i)}$ indicates a repulsive (attractive) interaction.  We quote results in units of $\kappa$.   The two-site model can be diagonalized and its spectrum directly determined from its eigenvalues.  We use the transfer matrix formalism \cite{Lee:2008fa} to obtain this spectrum \footnote{We compare the spectrum obtained from the transfer matrix (instead of direct diagonalization of the Hamiltonian) since it incorporates lattice discretization effects.  This enables a direct comparison with our lattice projection calculations.}.  With the spectrum in hand, we compare these results with those obtained from a stochastic lattice projection calculation \cite{Lee:2008fa} where the introduction of auxiliary fields via our transformation is needed.  A detailed description of these calculations is in preparation \cite{our-long-paper}.   In what follows, we give a succinct description.

The projection method extracts the lowest energy level $E$ in the spectrum of the system via
\begin{equation}\label{eqn:E}
E=-\lim_{\tau\to\infty}{\partial_\tau \log Z[\tau,\Psi_T]} \ ,
\end{equation}
where $Z$ is given by
\begin{align}
Z[\tau,\Psi_T] &\equiv\langle \Psi_T| e^{-\tau :\hat H:}|\Psi_T\rangle      \nonumber\\
&= \int \left(\prod_{x} d\phi_x P_N(\phi_x)\right) \K[\tau,\vec{\phi},c_j,\Psi_T],                     \label{eqn:Z}
\end{align}
$\Psi_T$ is an initial trial wavefunction, and $\vec\phi$ indicates the collection of all $\phi$ over $x$, all space (both sites) and time from 0 to $\tau$.

In the last step of the \eqref{eqn:Z} we have applied our transformation with its associated probability distribution $P_N(\phi)$ at each spacetime point and introduced $\K$ which is a functional of $\phi$ and depends explicitly on the coefficients $c_j$, the trial wavefunction $\Psi_T$, and the time separation $\tau$. The form of $P_N(\phi)$ depends on the order $N$ of the transformation that is applied.  For the work presented here, we consider the two extreme cases
\begin{align}
    N&=1        & P_1(\phi) &= \frac{1}{\sqrt{\pi}}\exp\left(-\phi^2\right)   \\
    N&=\infty   & P_\infty(\phi) &= \begin{cases}
        1/2 & |\phi| < 1 \\
        0 & \text{otherwise}
        \end{cases}
\end{align}
The case $N=1$ corresponds to the gaussian distribution of the original Hubbard-Stratonovich transformation (and thus only $\lambda^{(2)}$ is nonzero), whereas for $N=\infty$ we have uniform sampling and in principle all allowed many-body contact interactions.  In the $N=\infty$ case, there are in principle an infinite number of $c_j$---we set them all to zero for $j>3$.

As our trial wavefunction for multifermion systems $|\Psi_T\rangle$ we pick a direct product state of single-particle ground state wavefunctions, $|\psi_f\rangle = \frac{1}{\sqrt{2}}[a^\dag_{f,0}-a^\dag_{f,1}]|0\rangle$ for each flavor $f$.
The direct product structure allows us to write the functional $\K$ in \eqref{eqn:Z} as
\begin{equation}
    \K[\tau,\vec{\phi},c_j,\Psi_T]=\left(K^{-1}[\tau;\vec{\phi},c_j]\right)^\alpha
\end{equation}
where $\alpha$ is the number of fermion types and 
{\small\begin{align}
K^{-1}[\tau;\vec{\phi},c_j] &=\frac{1}{2}\left(
K^{-1}[0,\tau;0,0;\vec{\phi},c_j] -K^{-1}[0,\tau;1,0;\vec{\phi},c_j]\right.                             \nonumber\\
&\left.-K^{-1}[1,\tau;0,0;\vec{\phi},c_j]
+K^{-1}[1,\tau;1,0;\vec{\phi},c_j]\right)\ ,
\end{align}}
After introducing a time discretization $\tau = a_\tau t$, the $K$ matrix is given by
\begin{align}
K[i,t;i',t';\vec{\phi},c_j]\equiv   \delta_{i,i'}\delta_{t,t'}+a_\tau\delta_{\langle i,i'\rangle}\delta_{t,t'+1} \nonumber\\
                                    +\delta_{i,i'}\delta_{t,t'+1}\left(c_1\phi_{i,t}+c_2\phi_{i,t}^2+c_3\phi_{i,t}^3-1\right).
\end{align}
Here we have used the fact that we let $c_j=0$ for $j>3$.   We attach the indices $i$ and $t$ on the auxiliary field to indicate that there exists auxiliary fields for each spatial and temporal point.

Because we work with discretized time, we analyze the discretized version of eq.~\eqref{eqn:E},
\begin{equation}
E(\tau)\equiv-\frac{1}{a_\tau}\log\left(\frac{Z[\tau+a_\tau,\Psi_T]}{Z[\tau,\Psi_T]}\right)
\end{equation}
and search for constant plateaus at long times to numerically determine the $\tau\rightarrow\infty$ limit.

In Fig.~\ref{fig:spectrum} we show the numerical results of $E(\tau)/\kappa$ for combinations of $\lambda^{(M)}$ where $\lambda^{(2)}\in\{\pm1.74,0.0\}$ and $\lambda^{(3)}\in\{\pm 0.245,0.0\}$ for two and three-fermion systems.   For calculations with $\lambda^{(3)}=0.0$,  $\phi$ was sampled according to the HS distribution $P_1$ and all other calculations sampled according to $P_\infty$.  Note that the coefficients $c_j$ can be complex.  The solid lines correspond to the exact answers from diagonalization of the transfer matrix, and the dashed lines correspond to non-interacting energies.  We note that for all two-fermion calculations, both $N=1$ and $N=\infty$ calculations agree with each other, regardless of the value of $\lambda^{(3)}$, since the 3-body interaction plays no role.  Furthermore, with $\lambda^{(2)}=0.0$ but nonzero $\lambda^{(3)}$, the two-fermion system reproduces the non-interacting result.  In the three-fermion system, the effects of $\lambda^{(3)}$ are apparent and agree well with exact diagonalization.  A quantitative analysis will be given in \cite{our-long-paper}, as well as an analysis of signal-to-noise behavior for the different systems.

\emph{Discussion.}  The Hubbard-Stratonovitch transformation has been a useful tool for making analytic and numerical progress in problems of physical interest, and is a special case of the transformation described in this work.
Our transformation, which uses a self-interacting auxiliary field, allows for the direct inclusion of controllable many-body forces into numerical calculations.

Choosing a degree of auxiliary field self-interaction fixes the sampling distribution and limits the types of fermion/auxiliary-field interaction vertices.  These interactions in turn generate a slew of $n$-body forces.  Sampling a gaussian distribution for the auxiliary field recovers the original HS transformation, while uniform sampling in principle allows for the independent control of all possible $n$-body forces, while intermediate distributions yield correlated forces.

We have demonstrated that on a two-site model, different sampling techniques reproduce the exact results in a variety of cases, correctly handling all combinations of attractive, repulsive, and absent two-body forces with attractive and repulsive three-body forces.  While even the agreement of the one-body energies found with the different methods is nontrivial, the agreement of the two-body systems between methods and agreement of three-body systems with exact results indicate a reliable understanding and definite control over many-fermion interactions.

One drawback of our method is that some combinations of strengths of many-body forces are only achievable with complex $c_j$, opening the possibility of introducing a numerical sign problem \cite{Chen:2004rq}.  Because the $c$ coefficients appear nonlinearly in the higher-body $\lambda$ coefficients, there is not a unique set of $c$s that yield a particular set of $\lambda$s.

It would be interesting to consider sampling by more generic functions such as $P_N(\phi)=\exp(-\sum_n \alpha^{(n)}\phi^{n})$ with multiple tuneable $\phi$ self-interactions.  For field-theoretic applications, it may prove fruitful to understand how the renormalization of different $\lambda^{(M)}$ control renormalization of the parameters $c_j$.

We expect our transformation to be useful in a variety of physical systems where many-body interactions are important.  Numerical applications may include density functional theory approaches to nuclear physics \cite{Drut:2009ce}, the nonperturbative inclusion of multinucleon forces into NLEFT which might unlock precision characterizations of halo nuclei, the study of systems near the Efimov threshold such as cold atoms (see Refs.~\cite{Braaten:2006vd,Naidon:2016dpf} and references therein), systems at high density, and any other system where contact interactions need stochastic implementation.

\appendix
\section{Acknowledgements}
We thank A. Rokash, D. Lee and A. Cherman for insightful discussions.  This work was done in part through financial support from the Deutsche Forschungsgemeinschaft (Sino-German CRC 110).

\bibliography{references}

\end{document}